\documentclass[12pt]{article}
\usepackage{graphicx}
\usepackage{amssymb, amsmath, slashed}
\usepackage[utf8]{inputenc}

\newcommand\pubnumber{CIPANP2015-Deppisch}
\newcommand\pubdate{}

\def\ucl{Department of Physics and Astronomy\\
University College London, London WC1E 6BT, United Kingdom}

\def\Title#1{\begin{center} {\Large #1 } \end{center}}
\def\Author#1{\begin{center}{ \sc #1} \end{center}}
\def\Address#1{\begin{center}{ \it #1} \end{center}}

\newcommand\pubblock{\rightline{\begin{tabular}{l} \pubnumber\\
         \pubdate  \end{tabular}}}
\newenvironment{Abstract}{\begin{quotation}  }{\end{quotation}}
\newenvironment{Presented}{\begin{quotation} \begin{center} 
             PRESENTED AT\end{center}\bigskip 
      \begin{center}\begin{large}}{\end{large}\end{center} \end{quotation}}
\def\Acknowledgements{\bigskip  \bigskip \begin{center} \begin{large}
             \bf ACKNOWLEDGEMENTS \end{large}\end{center}}

\def\beq{\begin{equation}}
\def\eeq#1{\label{#1}\end{equation}}
\def\eeqn{\end{equation}}

\def\beqa{\begin{eqnarray}}
\def\eeqa#1{\label{#1}\end{eqnarray}}
\def\eeqan{\end{eqnarray}}

\let\bar=\overbar

\def\Dslash{\not{\hbox{\kern-4pt $D$}}}
\def\dslash{\not{\hbox{\kern-2pt $\del$}}}

\def\msb{{\bar{\ssstyle M \kern -1pt S}}}

\begin{document}
\begin{titlepage}
\pubblock

\vfill
\Title{Probing Leptonic Models at the LHC}
\vfill
\Author{Frank F. Deppisch}
\Address{\ucl}
\vfill
\begin{Abstract}
Models of neutrino mass generation provide well motivated scenarios of Beyond-the-Standard-Model physics. The synergy between low energy and high energy LHC searches facilitates an effective approach to rule out, constrain or ideally pinpoint such models. In this proceedings report, we provide a brief overview of scenarios where searches at the LHC can help determine the mechanism of light neutrino masses and potentially falsify baryogenesis mechanisms.
\end{Abstract}
\vfill
\begin{Presented}
Twelfth Conference on the Intersections of\\[0.1mm]Particle and Nuclear Physics (CIPANP 2015)\\[1.5mm]Vail Colorado, USA, May 19--24, 2015
\end{Presented}
\vfill
\end{titlepage}
\def\thefootnote{\fnsymbol{footnote}}
\setcounter{footnote}{0}

\section{Introduction}
With the discovery of the Higgs boson at the LHC and the determination of its couplings to fermions, we are tantalizingly close to verifying the mechanism of charged fermion mass generation. What will remain missing though is an understanding of the light neutrino masses. The observation of neutrino oscillations shows that neutrinos have finite masses and that individual lepton flavour is violated. Neutrinos are also usually considered to be Majorana particles, an assumption that facilitates an understanding of their small masses. It is natural to expect that the violation of the individual lepton flavours and, in the case of Majorana neutrinos, the total lepton number will show up in other contexts as well. This for example includes rare lepton flavour violating (LFV) decays of muons and taus and the total lepton number violating (LNV) neutrinoless double beta ($0\nu\beta\beta$) decay.

Quite generally, the possible violation of lepton flavour/number should be searched for at all energies that are experimentally accessible. This is because the observation of such processes would equally allow us a direct insight into the mechanism of neutrino mass generation. The most popular example is the so called seesaw mechanism (of type I) in which heavy right-handed Majorana neutrinos $N$ with masses $\gtrsim 10^{11}$~GeV are added to the Standard Model (SM). Their Yukawa coupling with the left-handed neutrinos induces the light Majorana masses of light neutrinos after electroweak (EW) symmetry breaking. This motivates the lightness of neutrinos through the breaking of lepton number symmetry at a very high scale~\cite{Minkowski:1977sc}.

Despite its popularity, the default type-I seesaw mechanism has two major phenomenological issues: (i) In the expected regime with $m_N \gtrsim 10^{11}$~GeV, the heavy neutrinos are far too heavy to be probed experimentally; (ii) The heavy neutrinos are sterile, i.e. gauge singlets and they only interact through a small mixing with light neutrinos. In this short proceedings report, we will briefly review two scenarios that instead include TeV scale and potentially non-singlet neutrinos and which can be probed at the LHC. In addition, we will comment on the general impact of the experimental observation of LNV on baryogenesis models.

\section{Inverse Seesaw}

In the standard type-I seesaw model with the (one generation) mass matrix for the left- and right-handed neutrino,
\begin{align}
\label{eq:seesaw}
		\begin{pmatrix}
				0   & m_D \\
				m_D & m_N
		\end{pmatrix},
\end{align}
the mass of the light neutrino $\nu$ and its mixing $\theta$ with the heavy neutrino $N$ is given by $m_\nu = - m_D^2 / m_N$ and $\theta = m_D / m_N = \sqrt{m_\nu / m_N}$, respectively. Here, $m_D$ is the neutrino Dirac mass, expected to be of the order of the EW scale, and $m_N$ is an LNV Majorana mass of the right-handed. For an observed light neutrino mass scale $m_\nu \approx 0.1$~eV this yields $\theta \approx 10^{-5}\sqrt{\text{GeV} / m_N}$. For a GeV to TeV scale heavy neutrino the mixing is rather small. This will be very different in the inverse seesaw scenario \cite{mohapatra:1986bd} described by the mass matrix
\begin{align}
\label{eq:invseesaw}
		\begin{pmatrix}
				0   & m_D   & 0    \\
				m_D & \mu_R & m_N  \\
				0   & m_N   & \mu_S
		\end{pmatrix},
\end{align}
similarly for the left-handed neutrino, the right-handed neutrino and an additional SM gauge singlet state $S$. Due to the presence of the small lepton number violating mass parameters $\mu_R$ and $\mu_S$, light neutrino masses are achievable for any $\theta = m_D / m_N$ \cite{gonzalezgarcia:1988rw}; in the simplest inverse scenario with $\mu_R = 0$ one has $\theta \approx 10^{-2} \sqrt{\text{keV} / \mu_S}$. The reason for this suppression can be understood as the two heavy neutrino states formed by $N$ and $S$ have opposite $CP$ parities and they combine to form quasi-Dirac neutrinos with a fractional mass splitting of order $\mu_S / m_N$. All lepton number violating observables, such as the light neutrino mass, will be suppressed by this small mass splitting.

\begin{figure}[t]
\centering
\includegraphics[clip,width=0.60\columnwidth]{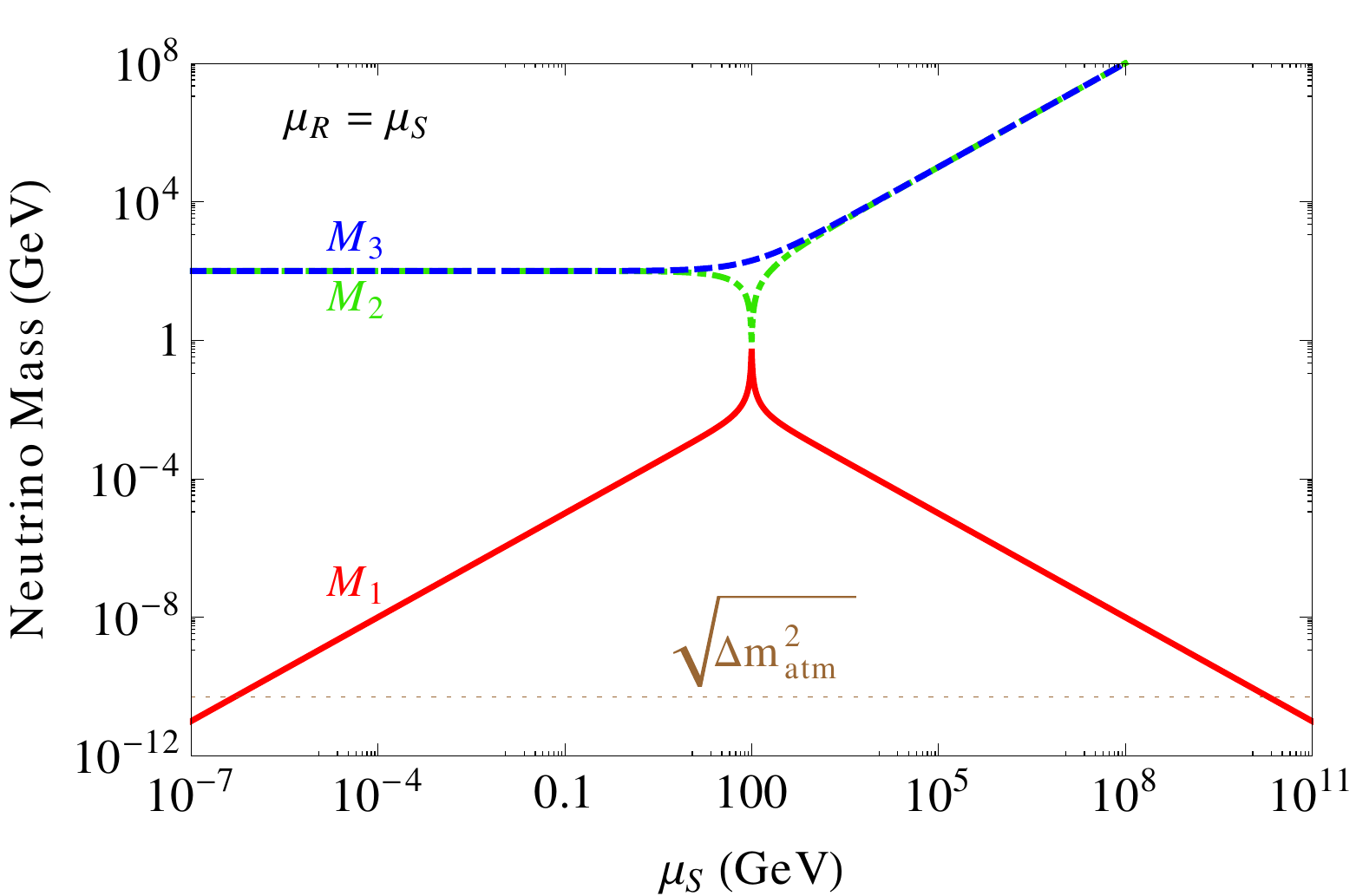}
\includegraphics[clip,width=0.39\columnwidth]{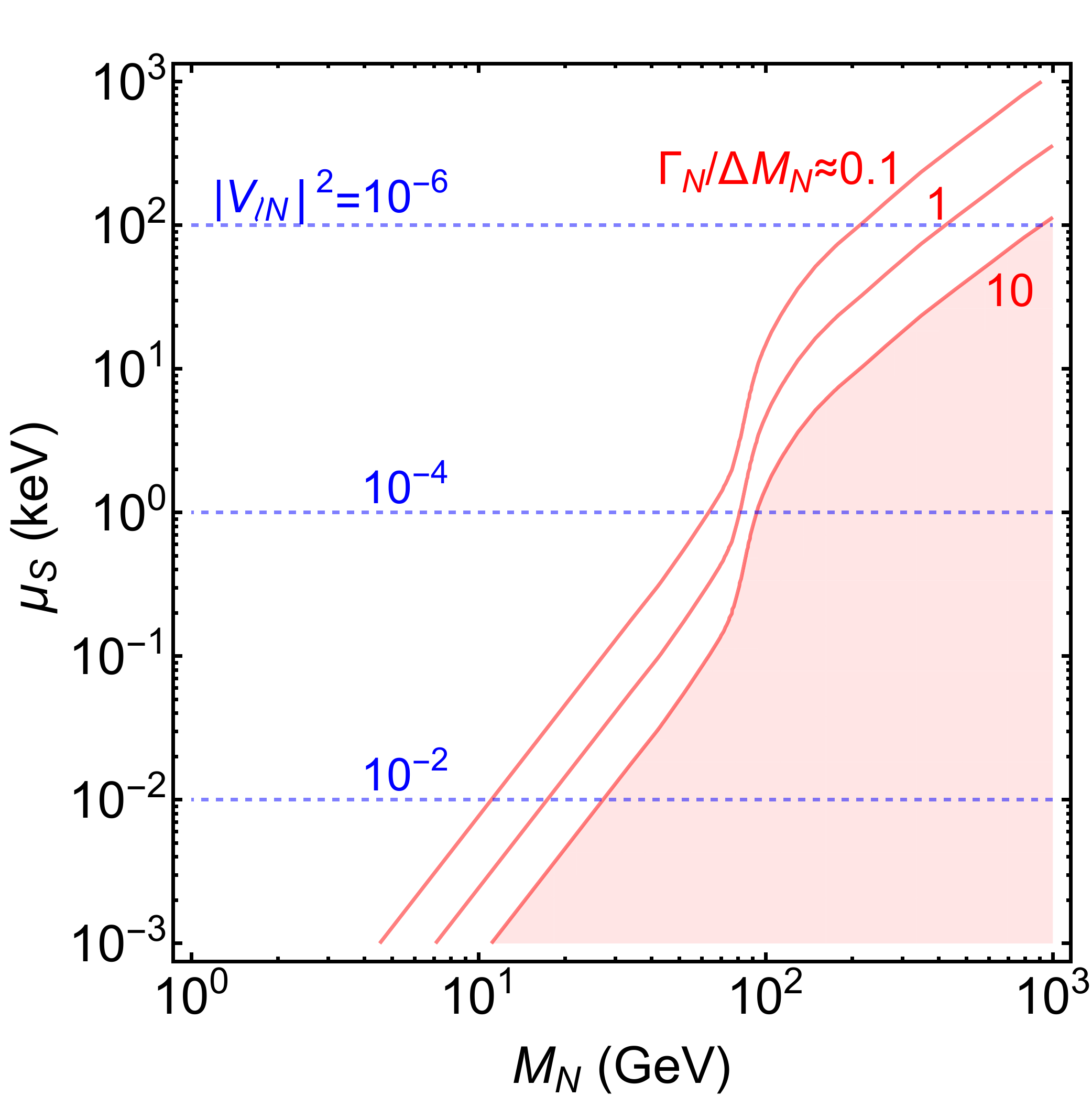}
\caption{Left: Spectrum of neutrino mass eigenstates as a function of the LNV parameter $\mu_S = \mu_R$ with $m_N = 100$~GeV and $m_D = 1$~GeV. The horizontal line denotes the desired light neutrino mass scale $m_\nu \approx 0.05$~eV for the lightest state. Right: Contours of the ratio of the average decay width of sterile heavy neutrinos to their mass splitting $\approx \mu_S$ in the inverse seesaw model. The LNV signal is expected to be unobservably small in the red shaded region with $\Gamma_N / \Delta m_N \gtrsim 10$. The horizontal blue lines denote contours of constant active-sterile mixing $|V_{lN}|^2$. Taken from \cite{Deppisch:2015qwa}.}
\label{fig:width}
\end{figure}
In order to see the transition between standard and inverse seesaw, we choose $\mu_S = \mu_R$ in eq.~\eqref{eq:invseesaw}. Fixing the other terms as $m_N = 100$~GeV and $m_D = 1$~GeV, Fig.~\ref{fig:width}~(left) shows that successful light neutrino mass generation occurs for $\mu \approx 10^{-6}$~GeV corresponding to the inverse seesaw and $\mu \approx 10^6$~GeV corresponding to the normal high-scale seesaw. The inverse case contains two heavy Majorana neutrinos with masses $m_N \pm \mu_S$ constituting a quasi-Dirac state, whereas for large $\mu \gg m_N$~GeV two heavy quasi-degenerate Majorana neutrinos with masses $\mu \pm m_N$ are formed.

\section{Heavy Sterile Neutrinos}

A large number of laboratory searches put constraints on the mixing between sterile and active neutrinos: For $m_N \ll 1$~MeV, sterile neutrinos are being probed in neutrino oscillation experiments. For pure Majorana sterile neutrinos, $0\nu\beta\beta$ searches provide stringent constraints on the mixing with electron neutrinos~\cite{Kovalenko:2009td, Atre:2009rg}, but these limits are considerably weakened for quasi-Dirac neutrinos such as found in the inverse seesaw mechanism discussed above. For $1~\text{MeV} \lesssim m_N \lesssim 1~\text{GeV}$, the active-sterile mixing is constrained by peak searches in leptonic decays of pions and kaons and in beam dump experiments. A more coherent overview of experimental searches for sterile neutrinos can be found in the recent review~\cite{Deppisch:2015qwa}.

Regarding LNV at high energy colliders, a general observation can be made in scenarios with approximately conserved lepton number like the above inverse seesaw mechanism: Like any LNV observable, the rate of an LNV process will be suppressed by the small mass splitting, but for on-shell resonant production of a heavy neutrino, the suppression is with respect to the neutrino width, $\Delta m_N / \Gamma_N$, rather than the absolute mass or the energy scale of the process. For $\Delta m_N \approx \Gamma_N$ it can be resonantly enhanced~\cite{Pilaftsis:1997dr}. The effect of the suppression is shown in Fig.~\ref{fig:width}~(right) giving contours of $\Gamma_N / \Delta m_N$ as a function of the inverse seesaw parameters $m_N$ and $\mu_S$. Within the shaded region the suppression would be too severe to expect an LNV observation. For $m_N \gtrsim 100$~GeV, one would require a small sterile-active mixing $|V_{\ell N}|^2 \lesssim 10^{-4}$ which suppresses the LNV rate as well.

In the specific context of the LHC, a Majorana heavy neutrino leads to a LNV signature with two same-sign leptons plus jets and no missing energy: $pp \to W^{(*)} \to N \ell^\pm \to \ell^\pm \ell^\pm jj$~\cite{Keung:1983uu, Pilaftsis:1991ug}. The CMS and ATLAS collaborations have performed direct searches for the production of heavy neutrinos limiting the mixing to active neutrinos $|V_{e(\mu) N}|^2 \lesssim 10^{-2}-10^{-1}$ for $m_N \lesssim 500$~GeV at $\sqrt s=8$~TeV~\cite{Khachatryan:2015gha}. During the ongoing run II of the LHC, the limits could be improved to apply to about a TeV. In addition to the basic $s$-channel production, it is also very worthwhile to consider other production modes and decay scenarios: Electroweak $t$-channel processes of the form $pp \to W^* \gamma^* \to N \ell^\pm jj$ can for example give a better sensitivity for higher $m_N$ values. Furthermore, searches for displaced vertices can considerably improve the sensitivity for heavy neutrinos lighter than the $W$ boson~\cite{Izaguirre:2015pga}.

\section{Left-Right Symmetry}

One of the simplest options to extend the above sterile neutrino scenario is an additional, broken $U(1)'$ gauge symmetry under which the heavy neutrinos are charged. Under favourable parameter conditions, heavy neutrinos can then be pair-produced abundantly and be probed even for very small mixing with the active neutrinos~\cite{Deppisch:2013cya}. Another popular option are left-right symmetric models (LRSMs); the minimal LRSM extends the SM gauge symmetry to $SU(2)_L\times SU(2)_R\times U(1)_{B-L}$~\cite{Pati:1974yy}. Leptons are assigned to doublets $L = (\nu, \ell)_L$ and $R = (N, \ell)_R$ under $SU(2)_L$ and $SU(2)_R$, respectively. The Higgs sector of the minimal LRSM consists of a bidoublet and two triplets ${\Delta}_{L,R}$. The VEV $v_R$ of the neutral component of ${\Delta}_R$ breaks the gauge symmetry $SU(2)_R \times U(1)_{B-L}$ to $U(1)_Y$ and gives masses to the RH gauge bosons $W_R$, $Z_R$ and the right-handed neutrinos $N$. LRSMs provide a simple ultraviolet complete seesaw mechanism with the key properties built in: The presence of right-handed neutrinos is a necessary ingredient and the LNV seesaw scale can be identified with the breaking scale of the $SU(2)_R$ symmetry.

\begin{figure}[t!]
\centering
\includegraphics[clip,width=0.43\textwidth]{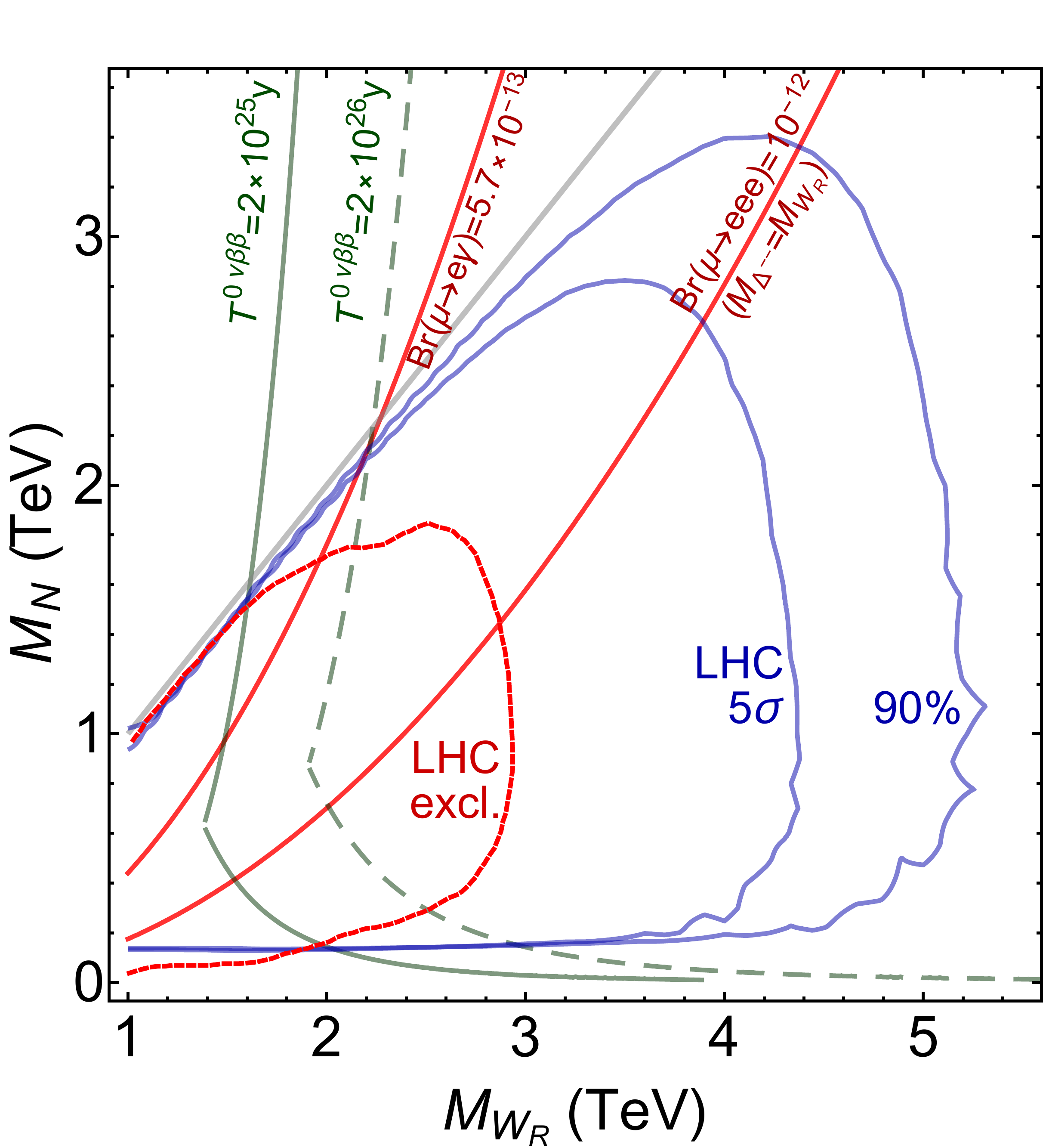}
\includegraphics[clip,width=0.45\textwidth]{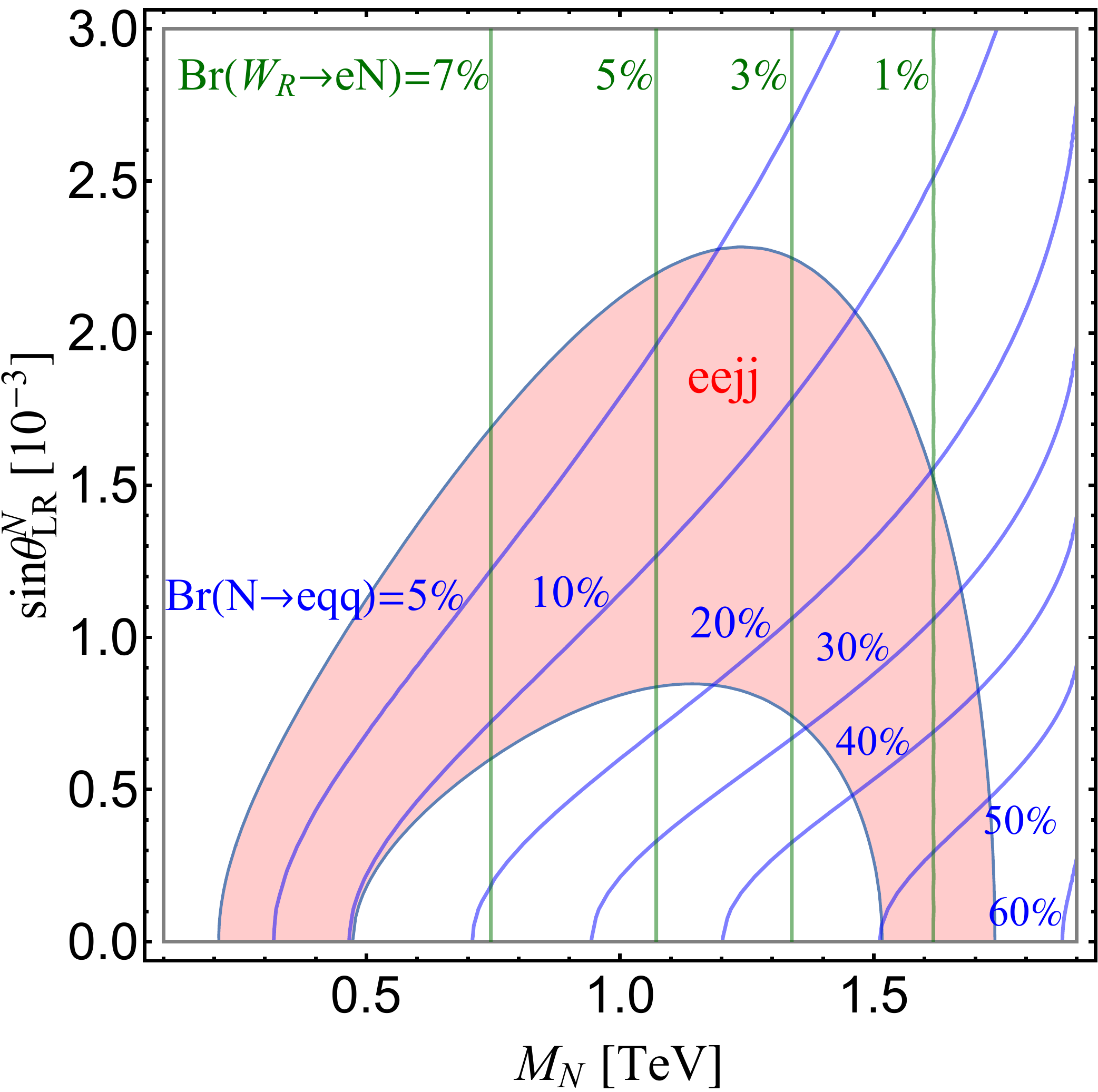}
\caption{Left: Sensitivity of the LHC to production and decay of heavy neutrinos via right-handed currents in the manifest LRSM. The solid blue contours give the signal significance of $5\sigma$ and 90\% at the LHC with 14~TeV and $\mathcal{L}=300\text{ fb}^{-1}$. The green and red contours show the sensitivity of $0\nu\beta\beta$ and LFV searches as denoted. Right: Fitting the CMS $eejj$ excess (red band) in the parameter plane of the heavy neutrino mass $M_N$ and the light-heavy mixing $\sin\theta^N_\text{LR}$. The other LRSM parameters are chosen in concordance with other excesses at 2~TeV as described in the text. Taken from \cite{Deppisch:2015cua}.}
\label{fig:LR}
\end{figure}
With regard to LHC searches, the right-handed current interactions in the LRSM can lead to a significant enhancement of the LNV signal. Even for negligible left-right neutrino mixing, heavy neutrinos can be directly produced via $s$-channel $W_R$ exchange~\cite{Keung:1983uu}. The potential to discover LFV and LNV at the LHC in this scenario has for example been analyzed in~\cite{Nemevsek:2011hz}. Fig.~\ref{fig:LR}~(left) compares the sensitivity of such searches with the sensitivity of $0\nu\beta\beta$ and low energy LFV experiments assuming equality of the $SU(2)_L$ and $SU(2)_R$ gauge couplings, $g_R = g_L$.

Both ATLAS and CMS have reported excesses in searches for dibosons, dijets and $e^\pm e^\mp jj$ using the LHC run I data, situated at invariant masses near 2~TeV. While far from being statistically significant, their coincidence at a resonant mass of 2~TeV is intriguing. Among several interpretations put forward in the literature, the excesses could be understood as a hint for $W_R$ production with $m_{W_R} \approx 2$~TeV, an $SU(2)_R$ gauge coupling $g_R \approx 0.6 g_L$ and a $W-W_R$ mixing of $\sin\theta_{LR}^W \approx 1.5\times 10^{-3}$~\cite{Deppisch:2014qpa}. Following through in this scenario, the CMS excess in $e^\pm e^\mp jj$~\cite{Khachatryan:2014dka} can be tentatively interpreted as the production of a heavy, quasi-Dirac neutrino as discussed above. This would allow to connect the LHC searches with neutrino physics, specifically the heavy neutrino mass and the strength of its mixing with the light neutrinos, cf.~Fig.~\ref{fig:LR}~(right).

\section{Lepton Number Violation and Baryogenesis}

The observed matter-antimatter asymmetry of the universe cannot be understood with SM physics. A large number of possible mechanisms to generate the observed asymmetry have been proposed in the literature. A particularly interesting scenario in our context is leptogenesis~\cite{fukugita:1986hr}. In its original formulation, the out-of-equilibrium and $CP$ violating decay of the heavy Majorana neutrinos in the type-I seesaw mechanism create a lepton asymmetry which is then converted into a baryon asymmetry through $(B+L)$ violating EW sphaleron processes~\cite{kuzmin:1985mm}.

\begin{figure}[t]
\centering
\includegraphics[clip,width=0.48\linewidth]{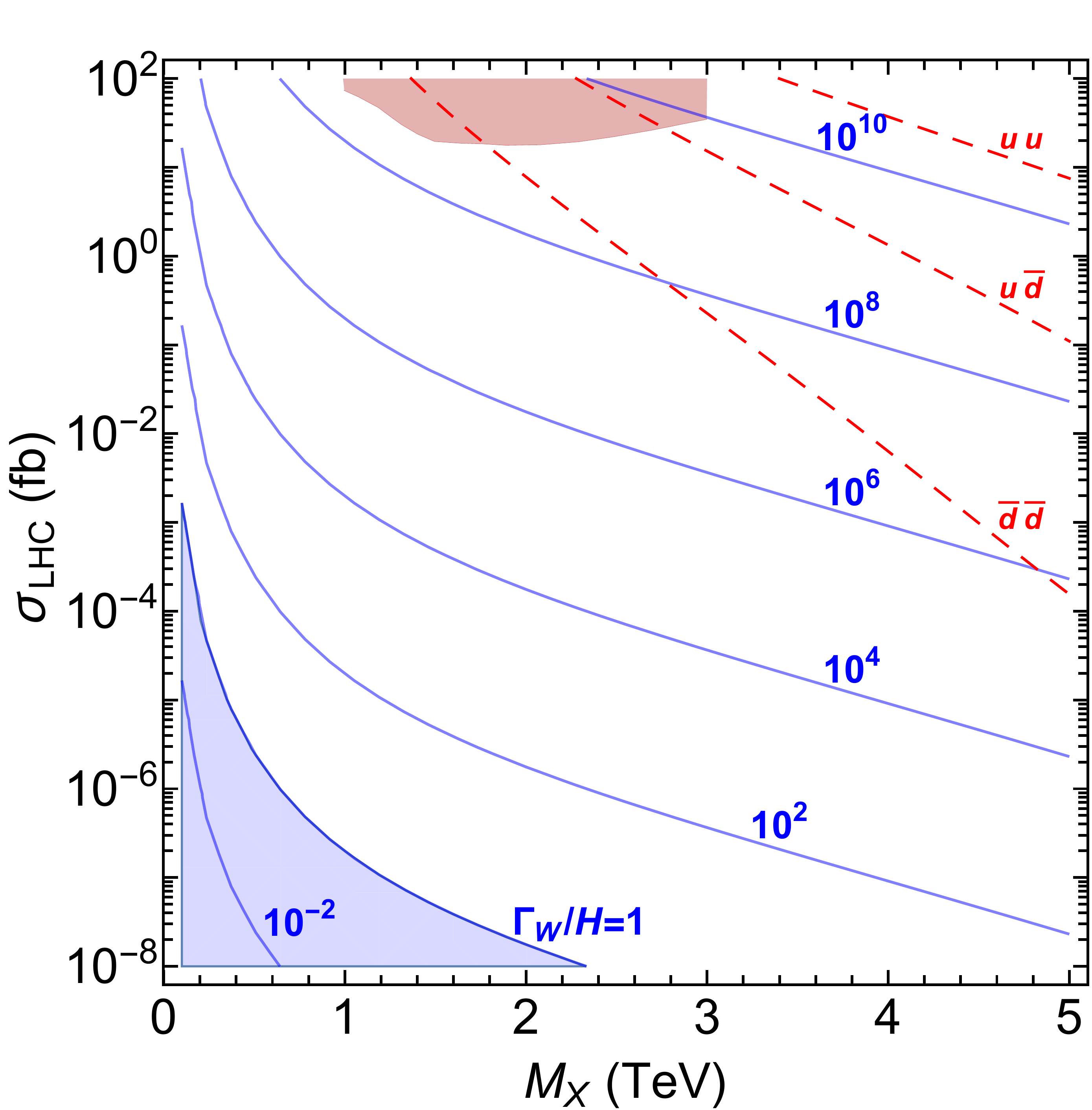}
\includegraphics[clip,width=0.45\linewidth]{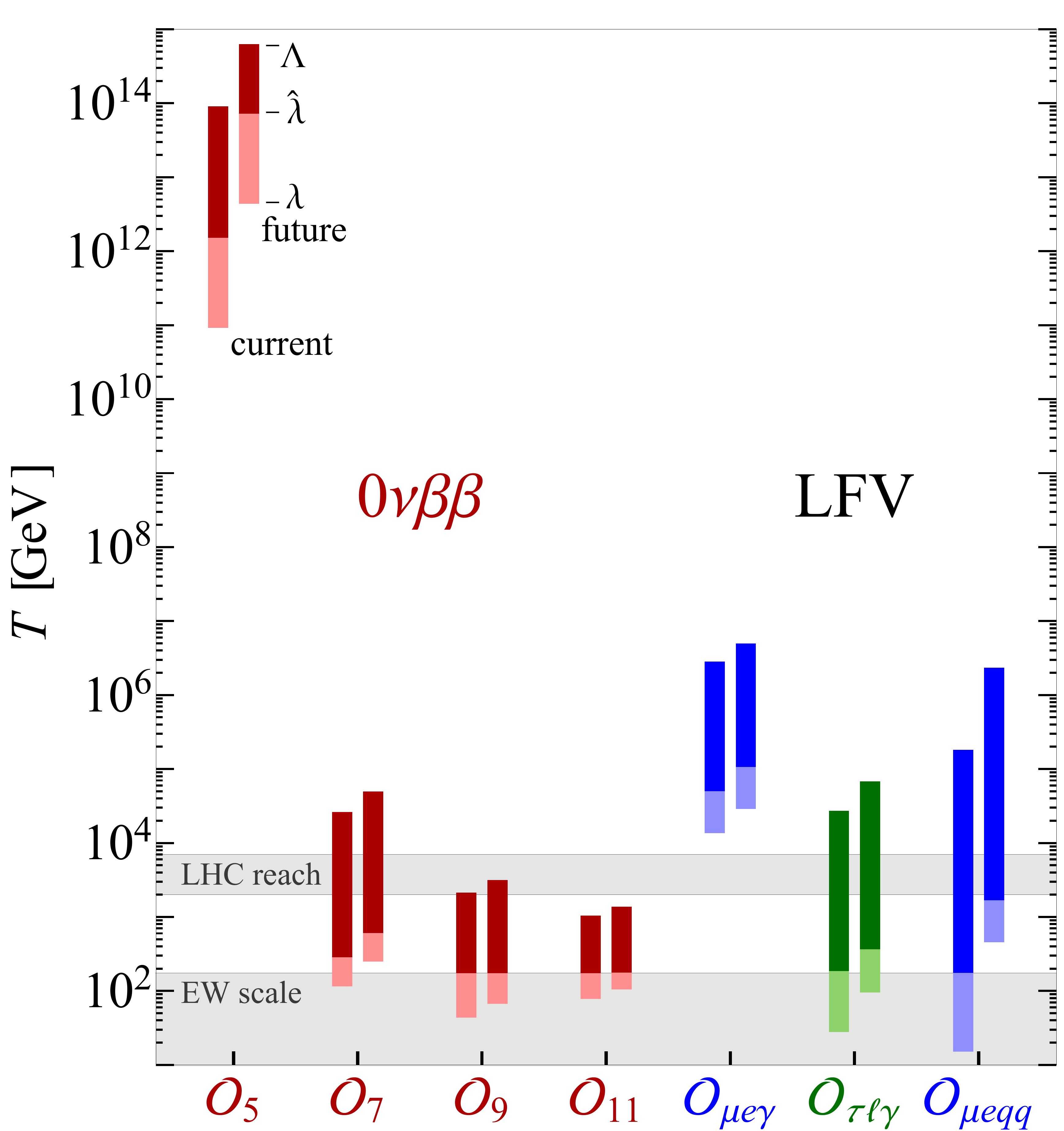}
\caption{Left: Lepton number washout rate in the early universe as a function of a hypothetically observed LNV scale $M_X$ and the LHC cross section $\sigma_\text{LHC}$ (solid blue contours). The red dashed curves denote typically expected cross sections for gauge strength interactions. Right: Temperature intervals where the given LNV and LFV operators are in equilibrium assuming that the corresponding process is observed at the current or future experimental sensitivity. Taken from \cite{Deppisch:2013jxa, Deppisch:2015yqa}.
}
\label{fig:decompositions}
\end{figure}
The presence of LNV is a crucial ingredient in leptogenesis. Furthermore, the observation of LNV would have important consequences on the viability of baryogenesis models in general; specifically, it is possible to falsify a large class of high-scale baryogenesis scenarios if LNV was observed at the LHC~\cite{Deppisch:2013jxa}. For example if a resonant LNV process with the signature $pp \to l^\pm l^\pm jj$ is observed, its LHC cross section $\sigma_\text{LHC}$ is related to the induced lepton asymmetry washout rate $\Gamma_W / H$ (relative to the expansion of the universe)~\cite{Deppisch:2013jxa}, 
\begin{align}
\label{eq:washout_factor_estimation}
  \log_{10}\frac{\Gamma_W}{H} &\gtrsim
  6.9 + 0.6\left( \frac{M_X}{\text{TeV}} - 1 \right) +
  \log_{10}\frac{\sigma_\text{LHC}}{\text{fb}}.
\end{align}
Here $M_X$ is the mass of the hypothetically observed resonance. If $\Gamma_W / H \gg 1$, the dilution of a primordial net lepton number density, understood to be produced in a baryogenesis mechanism at a higher scale, is highly effective and the lepton asymmetry would be washed out before it can be converted by sphaleron processes. This is illustrated in Fig.~\ref{fig:decompositions}~(left). Observation of LNV at the LHC would therefore rule out or strongly constrain baryogenesis scenarios above the scale $M_X$. 

A similar argument can be applied to non-standard mechanisms mediating $0\nu\beta\beta$ decay and low energy LFV processes~\cite{Deppisch:2015yqa}: if observed, the corresponding processes would be in equilibrium in certain temperature ranges. This is shown in Fig.~\ref{fig:decompositions}~(right) where the coloured bars denote the efficient equilibration temperatures assuming the relevant observable is seen at the current (left bar) or expected future (right bar) sensitivity. In the case of the 7,9,11-dimensional effective operators $\mathcal{O}_{7,9,11}$ mediating $0\nu\beta\beta$ decay, an electron lepton asymmetry present at higher energies would be washed out. Observation of LFV via 6-dimensional LFV operators at compatible scales would allow to extend the argument to other flavours than the electron.

\section{Conclusions}
Models of neutrino mass generation provide well motivated scenarios of BSM physics. The synergy between low energy and high intensity searches on the one hand and high energy LHC searches on the other are an effective approach to rule out, constrain or ideally pinpoint such models. In this report, we have briefly reviewed a few phenomenological scenarios where the LHC can help us to find out whether the light neutrino masses (and maybe the matter-antimatter asymmetry as well) are generated in a mechanism close to the electroweak scale or beyond.

\Acknowledgements
The author would like to thank the organizers of CIPANP2015 for the opportunity to contribute to the conference and the proceedings.

\end{document}